\documentclass[aps,prl,twocolumn,superscriptaddress]{revtex4-1}
\usepackage{bm}
\usepackage{graphicx}
\usepackage{color}
\usepackage{braket}
\usepackage{amsmath,amssymb,amsfonts,amsthm,mathtools}
\usepackage{enumerate}
\usepackage{enumitem}
\usepackage[colorlinks=true,linkcolor=blue,anchorcolor=red,citecolor=blue,urlcolor=blue]{hyperref}
\usepackage{titlesec}
\usepackage{float}
\usepackage{multirow}
\usepackage[title]{appendix}
\usepackage{pdfpages}
\usepackage{changes}
\usepackage{makecell}
\usepackage{diagbox}
\newcommand{\vect}[1]{\boldsymbol{#1}}

\makeatletter
\AtBeginDocument{\let\LS@rot\@undefined}
\makeatother

\makeatletter
\renewcommand\NAT@biblabelnum[1]{#1.}
\makeatother

\begin{document}

\title{Intrinsic in-plane magnetononlinear Hall effect in tilted Weyl semimetals}
\author{Longjun Xiang}
\affiliation{College of Physics and Optoelectronic Engineering, Shenzhen University, Shenzhen 518060, China}

\author{Jian Wang}
\email[]{jianwang@hku.hk}
\affiliation{College of Physics and Optoelectronic Engineering, Shenzhen University, Shenzhen 518060, China}
\affiliation{Department of Physics, University of Hong Kong, Pokfulam Road, Hong Kong, P. R. China}

\begin{abstract}
Armed with the extended semiclassical theory, we propose a Hall effect at $EB$ order, particularly
in Weyl semimetals (WSMs). We dub this effect the in-plane magnetononlinear Hall effect (IMHE) since
the Hall current and the driving electric and magnetic fields are confined in the same plane.
Similar to the intrinsic anomalous Hall effect, the IMHE features an intrinsic nature because
it arises from the field-induced anomalous velocity $\vect{E} \times \vect{\Omega}^B$,
where $\vect{\Omega}^B$ is the Berry curvature induced by the magnetic field through both minimal and Zeeman couplings.
Employing the low-energy effective Hamiltonian of WSMs,
we reveal that the tilt of the Weyl cone is the key to triggering this effect.
Notably, we find that the IMHE can survive even when the \textit{chiral anomaly} disappears
because $\vect{\Omega}^B$ (as the correction of the conventional Berry curvature)
does not contribute to the monopole charge. Furthermore,
we elucidate the interplay between minimal and Zeeman couplings for this effect.
Finally, the experimental strategy to detect the IMHE is discussed.
\end{abstract}

\maketitle

\noindent{\textit{\textcolor{blue}{Introduction.}}} ---
Capturing the intrinsic response of quantum matter
is one of the most important themes in modern condensed matter physics \cite{XGWen}.
For example, the quantum Hall effect \cite{Klitzing, TKNN, Haldane},
characterizing an intrinsic (linear) edge current response
in systems with broken time-reversal-symmetry, initialized the concept of topology
in condensed matter physics.
Recently, it was recognized that the intrinsic electric nonlinear Hall effect (ENHE) \cite{GaoY2014PRL}
with the response equation $j_a=\sigma_{abc}E_bE_c$
plays a pivotal role in detecting the reversal of the N\'eel vector
in antiferromagnetic spintronics \cite{GaoY2021PRL, XiaoC2021PRL}
as well as in probing the quantum metric of antiferromagnetic topological insulators \cite{SYXu2023, WBGao2023}.
However, ruled by the $\mathcal{T}$-odd and $\mathcal{P}$-odd characteristic of $\sigma_{abc}$,
the intrinsic ENHE can not be expected in quantum materials \cite{Nagaosa2017, JEMoore2017}
with time-reversal ($\mathcal{T}$) or inversion ($\mathcal{P}$) symmetries,
such as $\mathcal{T}$-symmetric or $\mathcal{P}$-symmetric Weyl semimetals (WSMs) 
\cite{XGWan, HMWeng, Hasan, X-Dai, Armitage}.

Along with the intrinsic ENHE arising from $\vect{E} \times \vect{\Omega}^E$,
the intrinsic magnetononlinear Hall effect (MHE) due to $\vect{E} \times \vect{\Omega}^B$
was also first proposed in Ref. [\onlinecite{GaoY2014PRL}] but received less attention,
especially for quantum materials,
where $\vect{\Omega}^E$ and $\vect{\Omega}^B$ stand for the Berry curvature corrections
induced by the electric field and the magnetic field, respectively.
Remarkably, the MHE conductivity $\sigma_{ab,c}$ defined by $j_a=\sigma_{ab,c}E_bB_c$ \cite{comma}
is a $\mathcal{P}$-even and $\mathcal{T}$-even (pseudo) tensor
and thereby can be used to probe the quantum materials, such as WSMs,
regardless its $\mathcal{P}$ and $\mathcal{T}$ symmetries.
Despite the extensive study of magnetotransport phenomena 
in WSMs especially under coplanar electromagnetic fields,
such as negative magnetoresistance \cite{C-Zhang, LMR, Arnold2016, ZrTe5, Sumiyoshi2016}
and the planar Hall effect
\cite{Nandy, D-Ma, P-Li, Kumar, F-Chen, Singha, Yang2019, Battilomo2021, Zyuzin2020, DXinplane0, Kundu2020},
which are intimately related to the \textit{chiral anomaly}
\cite{C-Zhang, chiralanomaly2, Berniverg2017},
the intrinsic MHE in WSMs under coplanar electromagnetic fields has not been discussed.

In addition, when it comes to the (in-plane) magnetic field,
the orbital contribution through the minimal coupling is focused on,
while the spin contribution through the Zeeman coupling is usually ignored
in the semiclassical treatment \cite{Berniverg2017, PALee}, 
although the in-plane Zeeman field is essential for
tailoring the topological properties of quantum materials including WSMs \cite{DXinplane0, DXinplane1, DXinplane2}.
Furthermore, besides the minimal coupling, the (in-plane) magnetic field through the Zeeman coupling
can also induce a Berry curvature $\vect{\Omega}^B$ \cite{XiaoC2022arxiv}.
However, how the interplay between both couplings
is manifested in the magnetotransport of WSMs remains elusive.

\begin{figure*}[t!]
\includegraphics[width=1.80\columnwidth]{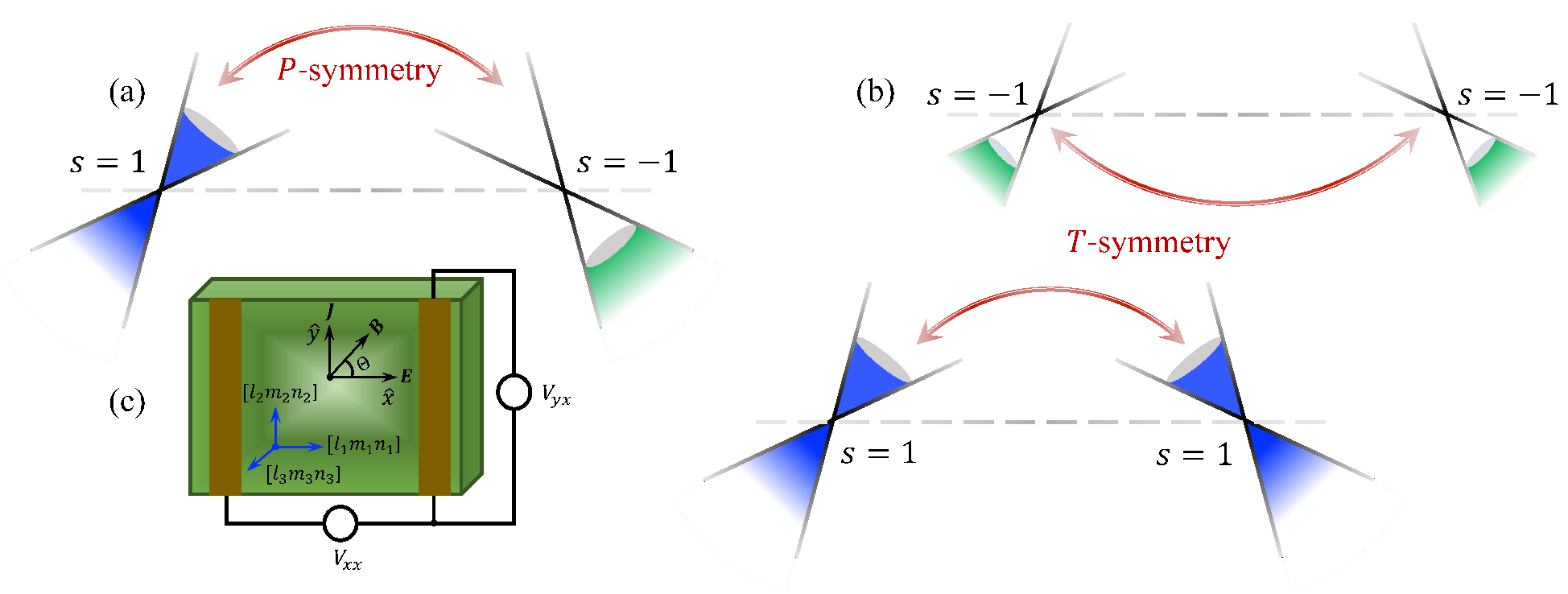}
\caption{
(a) A pair of $\mathcal{P}$-related WPs in magnetic WSMs.
(b) Two pairs of $\mathcal{T}$-related WPs in nonmagnetic WSMs.
For brevity, we suppress the momentum dependence of the chirality index $s$.
The blue and green shading between a pair of WPs with opposite chirality
illustrates the concept of \textit{chiral anomaly}: the number of particles with a given chirality is no longer
conserved when the electric and magnetic fields are applied in parallel: $\vect{E}//\vect{B}$.
(c) The schematic device to detect the intrinsic IMHE experimentally.
Here $[lmn]$ is the Miller index which can be along the principal axis of the crystal so that
$\vect{t}$ can have a nonzero $t_a$ component with $a=x, y, z$.
In addition, the in-plane configuration for the Hall current $\vect{J}$,
the electric field $\vect{E}$, and the magnetic field $\vect{B}$ is shown.}
\label{FIG1}
\end{figure*}

In this work, we investigate the intrinsic in-plane MHE (IMHE) \cite{inplane} for
magnetic and nonmagnetic WSMs with a type-I tilt senario,
based on the extended semiclassical theory \cite{GaoY2014PRL, XiaoC2022arxiv}.
We reveal that the field-induced anomalous velocity $\vect{E} \times \vect{\Omega}^B$
is the physical origin of this effect.
Particularly, employing the low-energy effective Hamiltonian of WSMs,
we analytically show that the tilt of WSMs plays a decisive role in triggering this effect,
where an out-of-plane tilt $t_z$ together with a further in-plane tilt $t_x$ or $t_y$
is necessary to observe this response.
In addition, we find that the IMHE can appear in WSMs even when the \textit{chiral anomaly}
is switched off by setting $\vect{E}\cdot\vect{B}=0$,
which is due to the fact
that $\vect{\Omega}^B$ (as the correction of the conventional Berry curvature $\vect{\Omega}$ \cite{D-Xiao})
does not contribute to the monopole charge.
Furthermore, for the minimal and Zeeman couplings,
we find that the IMHE conductivity shows a $\mu^{-1}$ and $\mu^{0}$ dependence
on the chemical potential $\mu$, respectively, and we illustrate that
the interplay between them is determined by the Fermi velocity $v_F$, the tilt, and the $g$-factor.
Finally, the experimental strategy to detect the IMHE in WSMs
is discussed, where the anti-symmetric property of the IMHE conductivity
can serve as a smoking gun to distinguish the possible competing effects.
Our work offers an intrinsic nonlinear Hall effect for diagnosing the WSMs.

\bigskip
\noindent \textcolor{blue}{\textit{Effective Hamiltonian for WSMs.}}---
The low-energy effective Hamiltonian for WSMs around a Weyl point (WP)
can be written as \cite{X-Dai, PALee, RHLi2021}:
\begin{align}
H_s = \vect{t} \cdot \vect{k} + s \vect{k}\cdot\vect{\sigma} + \mu_s, \label{WP1}
\end{align}
where $s = \pm 1$ is the chirality index, 
$\vect{k} \equiv \vect{q}-\vect{q}_i$ with $\vect{q}$ and $\vect{q}_i$ being the crystal momenta,
$\vect{\sigma}$ is the vector composed of the Pauli matrices;
$\vect{t}=(t_x, t_y, t_z)$ is the local tilt vector with
$|\vect{t}|<1$ corresponding to type-I WSMs \cite{X-Dai};
and $\mu_s$ denotes the energy shift relative to the charge neutral point of
the WP with chirality $s$. Note that we take $\hbar v_F=1$
in Eq. (\ref{WP1}), which can be easily restored in the final result by dimension analysis.

Usually, different WPs are related by symmetry \cite{PALee},
such as $\mathcal{P}$-symmetry or $\mathcal{T}$-symmetry.
For the former, because $\mathcal{P}\epsilon_n(\vect{q})=\epsilon_n(-\vect{q})$ with
$\epsilon_n$ being the global band dispersion for the $n$th band, we find that 
$\mathcal{P}\vect{k} \rightarrow -\vect{k} \Rightarrow \vect{t} \rightarrow -\vect{t}$ and
$\mathcal{P}s(\vect{q}_i)=-s(-\vect{q}_i)$ (note that $\mathcal{P}\vect{\sigma}=+\vect{\sigma}$),
which means that the $\mathcal{P}$-related WPs have an opposite tilt and chirality,
as illustrated in FIG. \ref{FIG1}(a).
Note that $\mathcal{P}s(\vect{q}_i)=-s(-\vect{q}_i)$
also implies that the minimal number of WPs for magnetic WSMs (particularly preserves $\mathcal{P}$ but breaks $\mathcal{T}$)
is two \cite{Nietwo}, as dictated by 
the Nielsen-Ninomiya fermion doubling theorem \cite{NoGo}, namely $\sum_i s(\vect{q}_i)=0$.
Similarly, for the latter, because $\mathcal{T}\epsilon_n(\vect{q}_i)=\epsilon_n(-\vect{q}_i)$
(ignore the spin quantum number for simplicity), we find that
$\mathcal{T}\vect{k} \rightarrow -\vect{k} \Rightarrow \mathcal{T}\vect{t} \rightarrow -\vect{t}$ and
$\mathcal{T}s(\vect{q}_i)=s(-\vect{q}_i)$ (note that $\mathcal{T}\vect{\sigma}=-\vect{\sigma}$),
which means that the $\mathcal{T}$-related WPs also have an opposite tilt but the same chirality.
As a result, the minimal number of WPs for nonmagnetic WSMs (particularly preserves $\mathcal{T}$ but breaks $\mathcal{P}$)
are four \cite{Hasanfour} to satisfy $\sum_{i}s(\vect{q}_i)=0$, as illustrated in FIG. \ref{FIG1}(b).
Finally, we note that the $\mathcal{P}$-related ($\mathcal{T}$-related)
two WPs have the same energy shift $\mu_{s(\vect{q}_i)} = \mu_{-s(-\vect{q}_i)}$
($\mu_{s(\vect{q}_i)} = \mu_{s(-\vect{q}_i)}$),
as indicated by the grey horizontal dashed lines in FIG. \ref{FIG1}(a) and \ref{FIG1}(b).

Considering that WPs always come in pairs
and can be tilted due to the reduced symmetries,
the resultant current contributed by different WPs is often determined by both the chirality and the tilt \cite{PALee}. 
However, sharply different from the magnetotransport behaviors previously discussed for WSMs,
the intrinsic IMHE proposed here is not dependent on the chirality index $s$.
As a consequence, we find that the two $\mathcal{P}$-related and $\mathcal{T}$-related
WPs in fact enjoy the same expression for the intrinsic IMHE conductivity,
as demonstrated below, which fundamentally originates from the
$\mathcal{T}$-even and $\mathcal{P}$-even characteristic of the intrinsic IMHE conductivity.

\begin{figure*}[tbp]
\includegraphics[width=1.80\columnwidth]{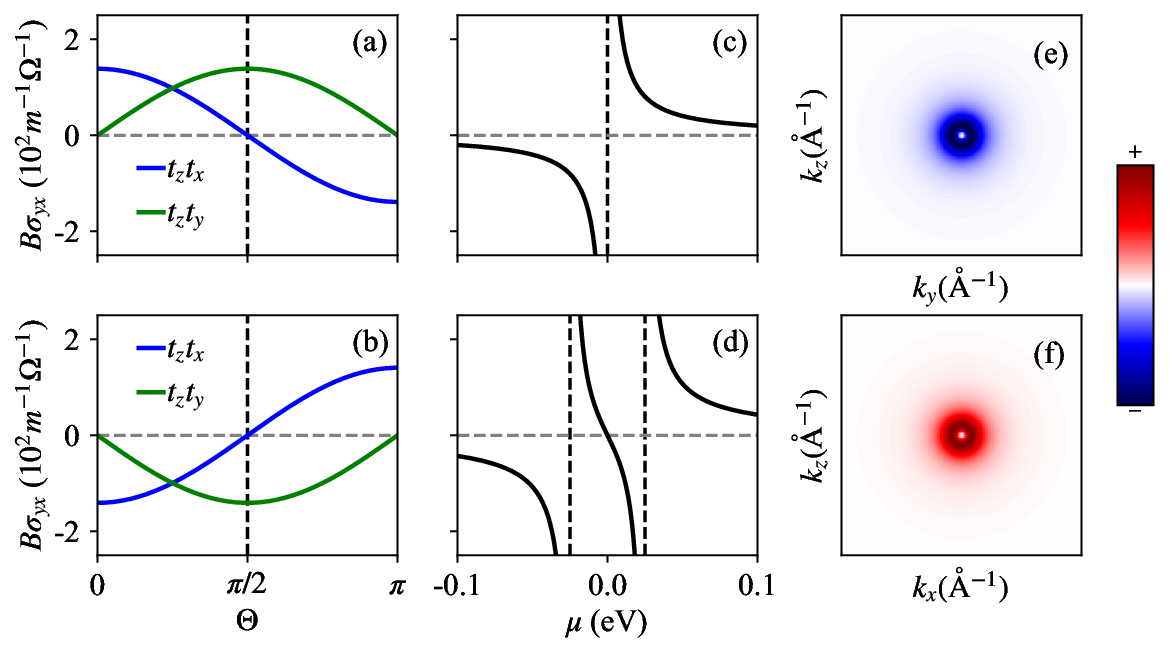}
\caption{The angular dependence of $B\sigma_{yx}$ for
(a) a pair of $\mathcal{P}$-related WPs and (b) two pairs of $\mathcal{T}$-related WPs that allowed by the tilt.
Parameters: $t_a=0.7$, $v_F=10^5 \mathrm{m/s}$, $B=10 \text{T}$, $\mu=0.0145 \mathrm{eV}$,
and $\mu_s=s25 \mathrm{meV}$ (only for $\mathcal{T}$-related WPs).
The chemical potential dependence of $B\sigma_{yx}$ for
(c) a pair of $\mathcal{P}$-related WPs and (d) two pairs of $\mathcal{T}$-related WPs that allowed by tilt,
the vertical dashed lines indicate the position of $\mu_s$.
The $\vect{k}$-resolved AOPs (for conduction band) that contribute to the intrinsic IMHE for
(e) $\vect{t}=(t_x,0,t_z)$ with $\Theta=0$ as well as $k_x=0.01 \mathrm{\mathring{A}}$
and (f) $\vect{t}=(0, t_y, t_z)$ with $\Theta=\pi/2$ as well as $k_y=0.01 \mathrm{\mathring{A}}$, respectively.}
\label{FIG2}
\end{figure*}
 
\bigskip
\noindent \textcolor{blue}{\textit{Extended semiclassical theory.}}---
The extended semiclassical equation of motion under electromagnetic fields
is given by \cite{GaoY2014PRL} ($\hbar=e=1$) :
\begin{equation}
\dot{\vect{r}} = \bar{\vect{v}}-\dot{\vect{k}} \times \bar{\vect{\Omega}},
\quad
\dot{\vect{k}} = -\vect{E}-\dot{\vect{r}} \times \vect{B},
\label{eq1}
\end{equation}
where $\bar{\vect{v}}=\vect{\nabla}_{\vect{k}} \bar{\epsilon}(\vect{k})$
is the group velocity, with $\bar{\epsilon}(\vect{k})$ being the band energy 
accurate up to the second order of the electromagnetic fields \cite{GaoY2014PRL},
and $\bar{\vect{\Omega}} = \vect{\Omega}+\vect{\Omega}^{(1)}$
with $\vect{\Omega}$ being the conventional Berry curvature \cite{D-Xiao}
and $\vect{\Omega}^{(1)}$ being the first-order field-induced one \cite{GaoY2014PRL}.
To be specific, $\vect{\Omega}^{(1)} \equiv \vect{\Omega}^{B} + \vect{\Omega}^{E}$
with $\vect{\Omega}^{B} \equiv \vect{\nabla}_{\vect{k}} \times \vect{\mathcal{A}}^{B}$
($\vect{\Omega}^{E} \equiv \vect{\nabla}_{\vect{k}} \times \vect{\mathcal{A}}^{E}$ )
the Berry curvature induced by $\vect{B}$ ($\vect{E}$),
where $\vect{\mathcal{A}}^{B} (\vect{\mathcal{A}}^E)$ is the $U(1)$ gauge-invariant
\textit{positional shift} \cite{GaoY2014PRL}.

In this work, we are interested in the intrinsic IMHE in WSMs
and hence we will focus on $\vect{\Omega}^B$ for two-band systems.
Note that $\vect{\Omega}^B$ and $\vect{\mathcal{A}}^B$ include
both the orbital and spin contributions through the minimal
and Zeeman couplings \cite{GaoY2014PRL, XiaoC2022arxiv}, respectively.
Particularly, by writing $\mathcal{A}_b^B \equiv B_a\mathcal{F}_{ab} \equiv B_a (\mathcal{F}^O_{ab} + \mathcal{F}^S_{ab})$
\cite{Esum},
with $\mathcal{F}^O_{ab}$ and $\mathcal{F}^S_{ab}$ being
the anomalous orbital polarizability (AOP) and the anomalous spin polarizability (ASP) \cite{GaoY2014PRL, XiaoC2022arxiv}, respectively,
for two-band systems we find
\begin{align}
\mathcal{F}^O_{ab}
=
-
\dfrac{\epsilon_{acd}(v_c^n+v_c^m)g_{db}^{nm}}{\epsilon_n-\epsilon_m}
-
\dfrac{\epsilon_{acd}\partial_c g^{nm}_{db}}{2}
\quad (m \neq n)
,
\label{orbital}
\\
\mathcal{F}^S_{ab} 
= - 2 \text{Re} \dfrac{\mathcal{M}_a^{S,nm}  \mathcal{A}_b^{mn} }{\epsilon_n-\epsilon_m}
\quad (m \neq n),
\label{spin}
\end{align}
where $m$ and $n$ belong to $\{+,-\}$, with $+ (-)$ being the conduction (valence) band;
$\partial_c=\partial/\partial k_c$;
$\epsilon_n$ stands for the band energy of the $n$th band;
$\mathcal{A}_b^{mn}$ is the interband Berry connection;
and $\epsilon_{abc}$ is the Levi-Civita symbol.
In addition, $g_{db}^{nm}=2\text{Re}\left[\mathcal{A}_d^{nm} \mathcal{A}_b^{mn} \right]$ is the quantum metric,
which measures the distance between the neighboring Bloch states \cite{SYXu2023, WBGao2023},
and $\mathcal{M}^{S,mn}_{b}=-g\mu_Bs_b^{mn}$ is the interband spin magnetic moments, 
where $g$ is the $g$-factor for spin, $\mu_B$ is the Bohr magneton,
and $s_b^{mn}$ is the interband matrix elements of spin operator.

Next, by solving Eq. (\ref{eq1}), we obtain
$
\dot{ \vect{r} } = \mathcal{D}^{-1} [ \bar{ \vect{v}} + \vect{E} \times \bar{ \vect{\Omega}} +
\vect{B} ( \bar{\vect{v}} \cdot \bar{\vect{\Omega}}) ],
$
where $\mathcal{D}=1+\vect{B} \cdot \bar{\vect{\Omega}}$ is the phase space factor.
Furthermore, by substituting this semiclassical velocity $\dot{\vect{r}}$
into the definition of charge current density \cite{D-Xiao},
$\vect{j} \equiv \int_{\vect{k}} \mathcal{D} \bar{f}_k \dot{\vect{r}}$,
we arrive at 
$
\vect{j}
=
\int_{\vect{k}} \bar{f}_k
[ \bar{ \vect{v}} + \vect{E} \times \bar{\vect{\Omega}}
+ \vect{B} (\bar{ \vect{v}} \cdot \bar{\vect{\Omega}})],
$
where $\int_{\vect{k}} \equiv \sum_s \int d\vect{k}/(2\pi)^d$,
with $d$ being the spatial dimension, and
$\bar{f}_k = f_k(\bar{\epsilon}_k)$ 
is the nonequilibrium Fermi distribution function that considers
the energy correction under magnetic field.
Importantly, from the anomalous velocity $\vect{E} \times \vect{\Omega}^B$,
the intrinsic IMHE conductivity is derived as
\cite{GaoY2014PRL, XiaoC2022arxiv}
\begin{align}
\sigma_{ab,c}
&=
\int_{\vect{k}} f' \left ( v_a \mathcal{F}_{cb} - v_b \mathcal{F}_{ca} \right) 
\label{cond11},
\end{align}
where the integration by parts is used and $\mathcal{F}_{cb}$ can either be $\mathcal{F}^O_{cb}$ or $\mathcal{F}^S_{cb}$.
Due to the presence of $f'=\partial f/\partial \epsilon$,
where $f$ is the equilibrium Fermi distribution without energy correction,
we conclude that $\sigma_{ab,c}$ features the Fermi surface property \cite{D-Xiao},
the same as the intrinsic ENHE \cite{GaoY2021PRL, XiaoC2021PRL}.
Note that in Eq.(\ref{cond11}) we have dropped a term contributed by
the conventional anomalous velocity $\vect{E}\times\vect{\Omega}$
in concert with the wavepacket energy correction $\vect{B}\cdot\vect{m}$
($\vect{m}$ is the orbital and spin magnetic momenta) \cite{GaoY2014PRL, XiaoC2022arxiv}
since it does not contribute to the intrinsic IMHE for type-I WSMs (see Appendix C).

We emphasize that $\sigma_{ab,c}$ is driven by the momentum-space
Lorentz force $\vect{E} \times \vect{\Omega}^B$,
which dictates that $\sigma_{ab,c}$ is an intrinsic and anti-symmetric tensor,
namely, $\sigma_{ab,c}=-\sigma_{ba,c}$, while $B_c$ remains unchanged.
This is different from the planar Hall effect \cite{Nandy} in WSMs caused by the \textit{chiral anomaly},
which corresponds to an extrinsic and symmetric tensor.
In addition, we note that the intrinsic IMHE is also different from the ordinary Hall effect,
where the applied magnetic field is perpendicular to the plane formed by the Hall current
and the applied electric field. Finally, we wish to mention that 
the chiral velocity $\vect{B}(\bar{\vect{v}}\cdot\bar{\vect{\Omega}})$
in the current expression does not contribute an intrinsic current at $EB$ order (see Appendix D).
However, the dispersive velocity due to the energy correction
at $EB$ order \cite{XiaoCPRB2021}
can make a similar contribution to the IMHE from $\vect{E} \times \vect{\Omega}^B$,
as discussed in Appendix E.
To close this section, we summarize that Eqs. (\ref{orbital})-(\ref{cond11})
are the main equations employed to explore the intrinsic IMHE in WSMs.

\begin{figure*}[tbp]
\includegraphics[width=1.8\columnwidth]{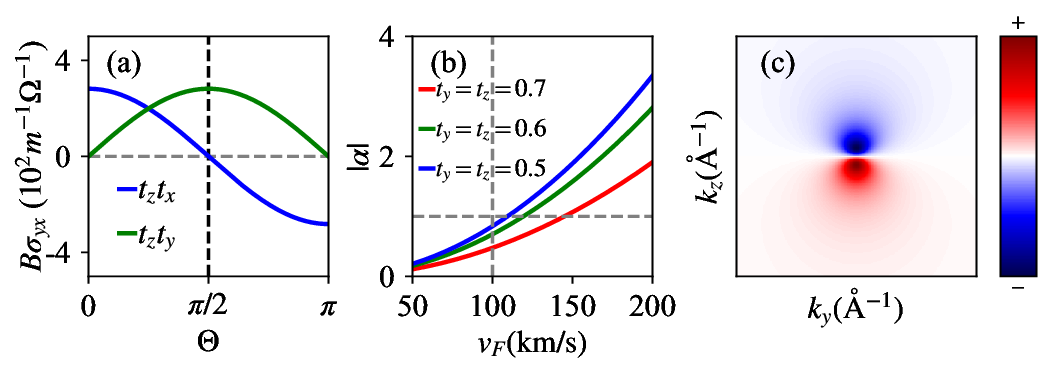}
\caption{
(a) The angular dependence of $B\sigma_{yx}$ due to the Zeeman coupling
for a pair of $\mathcal{P}$-related or $\mathcal{T}$-related WPs. 
Parameters: $v_F=10^5 \mathrm{m/s}$, $g=11.6$ \cite{DXinplane2}, and $t_a=0.7$.
(b) The orbital versus spin contribution for a fixed chemical potential $\mu=0.015 \mathrm{eV}$,
see Eq.(\ref{ratio}) for the definition of $\alpha$.
Here the gray vertical dashed line highlights the critical Fermi velocity, below which the spin contribution 
can be comparable to the orbital contribution.
(c) The $\vect{k}$-resolved ASP (for the conduction band) that contribute to the intrinsic IMHE.}
\label{FIG3}
\end{figure*}

\bigskip
\noindent{\textcolor{blue}{\textit{Intrinsic IMHE from the minimal coupling.}}} ---
We first consider the orbital contribution from the minimal coupling. Particularly,
for Eq.(\ref{WP1}), it is easy to show that $\epsilon^{\pm}=\vect{t}\cdot\vect{k}\pm k + \mu_s$,
where $k^2=k_x^2+k_y^2+k_z^2$,
and hence we have
\begin{align}
\vect{v}^\pm = \left( t_x \pm \hat{k}_x, t_y \pm \hat{k}_y, t_z \pm \hat{k}_z \right),
\label{velocity}
\end{align}
where $\hat{k}_a \equiv k_a/k$. In addition, the AOPs are given by
\begin{align}
\mathcal{F}^O_{x}
&=
\dfrac{
\mp
\alpha_{11} \cos\Theta
\mp
\alpha_{12}
\sin\Theta
}{4k^3},
\label{FOx}
\\
\mathcal{F}^O_{y}
&=
\dfrac{
\pm 
\alpha_{21}\sin\Theta 
\pm 
\alpha_{22} \cos\Theta
}{4k^3},
\label{FOy}
\end{align}
where $\mathcal{F}_{a}^O \equiv B_b\mathcal{F}^O_{ba}/B$ and we assume $\vect{B}=B(\cos\Theta,\sin\Theta,0)$,
as shown in FIG. \ref{FIG1}(c). 
Here
$\alpha_{11}=(t_z\hat{k}_y-t_y\hat{k}_z)\hat{k}_x$,
$\alpha_{12}=t_x\hat{k}_x\hat{k}_z+t_z (\hat{k}_y^2+\hat{k}_z^2 )$,
$\alpha_{21}=(t_z\hat{k}_x-t_x\hat{k}_z)\hat{k}_y$,
and
$\alpha_{22}=t_y\hat{k}_y\hat{k}_z+t_z(\hat{k}_{x}^2+\hat{k}_z^2)$.
Note that we dropped the second term of Eq. (\ref{orbital})
because it does not contribute to the intrinsic IMHE for type-I WSMs (see Appendix B).
In addition, at zero temperature we find \cite{RHLi2021, Das2019}
\begin{align}
f'_{\pm}=\delta (\epsilon^{\pm}-\mu)=\delta(k-\Delta \mu_s/\beta^\pm)/|\beta^\pm|,
\label{fermisurf}
\end{align}
where $\mu$ denotes the chemical potential, $\Delta\mu_s=\mu-\mu_s$, and
$\beta^{\pm} \equiv t_x\sin\theta\cos\phi+t_y\sin\theta\sin\phi+t_z\cos\theta \pm 1$
in spherical $\vect{k}$-space, namely 
$\vect{k}=k(\sin\theta\cos\phi,\sin\theta\sin\phi,\cos\theta)$.
Note that Eq. (\ref{fermisurf}) requires $k=\Delta\mu_s/\beta^{\pm}>0$,
which in fact is satisfied automatically for type-I WSMs
because $\beta^+>0$ and $\beta^-<0$ for $\mu>0$ and $\mu<0$, respectively,
where $\mu$ can penetrate only either the conduction band or the valence band.
Consequently, by substituting Eqs. (\ref{velocity})-(\ref{fermisurf}) into
Eq. (\ref{cond11}), we find (see Appendix A)
\begin{align}
\sigma^{s}_{yx} 
&=
\dfrac{1}{12\pi^2\Delta\mu_s} 
\left( \dfrac{e^3 v_F}{\hbar} \right)
t_z(t_x\cos\Theta+t_y\sin\Theta),
\label{result11}
\end{align}
where $\sigma^s_{yx} = \sigma^s_{yx,x}+\sigma^{s}_{yx,y}$ 
by defining $j_y^s \equiv \sigma_{yx}^sE_xB$ and $e, \hbar$, and $v_F$
are restored by dimension analysis.
Equation (\ref{result11}) is free of the chirality index $s$ but
quadratically depends on the tilt direction and hence the contribution from
an opposite tilt (regardless of its chirality) can be simply duplicated.
For a pair of $\mathcal{P}$-related WPs, by adding them we find:
\begin{align}
\sigma_{yx} = 
\dfrac{1}{6\pi^2 \mu}
\left( \dfrac{e^3 v_F}{\hbar} \right)
t_z(t_x\cos\Theta+t_y\sin\Theta),
\label{result12}
\end{align}
where we set $\mu_s=\mu_{-s}=0$;
for two pairs of $\mathcal{T}$-related WPs, by adding them we find
\begin{align}
\sigma_{yx} = 
\dfrac{1}{6\pi^2}
\dfrac{1}{\Delta\mu}
\left( \dfrac{e^3 v_F}{\hbar} \right)
t_z(t_x\cos\Theta+t_y\sin\Theta),
\label{result13}
\end{align}
where $1/\Delta\mu \equiv 1/\Delta \mu_{+} + 1/\Delta \mu_{-}$
with $\Delta \mu_{\pm}=\mu-\mu_{\pm}$.

From Eqs. (\ref{result12}) and (\ref{result13}),
it is easily found that the tilt vector $\vect{t}$
must have a nonzero projection on the $t_x$-$t_z$/$t_y$-$t_z$ plane
to capture the intrinsic IMHE \cite{Sugang}. Particularly, for the tilt-allowed IMHE responses,
the angular dependence of $B\sigma_{yx}$ for $\mathcal{P}$-related and $\mathcal{T}$-related
WPs is displayed in FIG. \ref{FIG2}(a) and FIG. \ref{FIG2}(b), respectively.
Interestingly, for $\vect{t}=(0, t_y, t_z)$, we find that
the IMHE response in both situations is nonzero even when $\Theta=\pi/2$,
namely when the \textit{chiral anomaly} is switched off.
In addition, in FIG.(\ref{FIG2}c) and FIG.(\ref{FIG2}d), we present
their chemical potential dependence of $B\sigma_{yx}$ for $t_zt_y$ tilt with $\Theta=\pi/2$,
from which we find that this response becomes significant when $\mu$
is close to the charge neutral point $\mu_s$, which is a small-gap effect \cite{XiaoC2022arxiv}.
Furthermore, to highlight the band geometric origin of the intrinsic IMHE, in FIGs. \ref{FIG2}(e) and \ref{FIG2}(f)
we show the $\vect{k}$-resolved AOPs that contributes to the intrinsic IMHE
for $\vect{t} = (t_x, 0, t_z)$ and $\vect{t} = (0, t_y, t_z)$, respectively.
Interestingly, although the AOPs display a monopole landscape, they do not, in fact, modify the
monopole charge arising from the conventional Berry curvature $\vect{\Omega}$ of WSMs (see Appendix F).
Finally, we note that $\sigma_{yx}$ shows a period of $2\pi$
due to the linear dependence on the magnetic field \cite{twopi}.

\bigskip
\noindent{\textcolor{blue}{\textit{Intrinsic IMHE from the Zeeman coupling.}}} ---
Next we consider the spin contribution through the Zeeman coupling.
Particularly, for Eq. (\ref{WP1}), we find that
the ASPs due to Zeeman coupling are given by:
\begin{align}
\mathcal{F}^S_x = \mp g \mu_B \dfrac{k_z}{2k^3} \sin\Theta,
\quad
\mathcal{F}^S_y = \pm g \mu_B \dfrac{k_z}{2k^3} \cos\Theta,
\label{FSxy}
\end{align}
which differ by a factor of $1/k$ compared to $\mathcal{F}^O_{x/y}$.
Performing a similar calculation, the intrinsic IMHE conductivity for two 
$\mathcal{P}$-related or $\mathcal{T}$-related WPs can be evaluated as (see Appendix A)
\begin{align}
\sigma_{yx} = \dfrac{g \mu_B}{8\pi^3} \left( \dfrac{e^2}{\hbar^2 v_F} \right) \left(C_1 \cos\Theta+C_2 \sin\Theta \right),
\label{spinsigmayx}
\end{align}
where $C_i \equiv \pm \int_0^{\pi}\int_0^{2\pi} \lambda_i^\pm (\vect{t};\theta,\phi)d\theta d\phi$
is a dimensionless constant [see Eqs. (\ref{lambda1}) and (\ref{lambda2}) for $\lambda_i$].
Interestingly, we find that the IMHE conductivity due to ASP
is independent of $\mu$ and hence the IMHE from Zeeman coupling is no longer a small-gap effect.
Additionally, to have a nonzero $C_1/C_2$, we also require at least that
$\vect{t}=(0,t_y,t_z)$/$\vect{t}=(t_x,0,t_z)$, as shown in FIG. \ref{FIG3}(a).
Similar to the orbital contribution, we find that the spin contribution
can also survive when the \textit{chiral anomaly} vanishes.

Importantly, we note that the spin contribution is on the same order as
the orbital contribution near $\mu = 0.015 \mathrm{eV}$ with $v_F = 10^5 \mathrm{m}/s$,
$t_a=0.7$, and $g = 11$ \cite{DXinplane2},
as can be seen by comparing FIG. \ref{FIG2}(a) with FIG. \ref{FIG3}(a).
In fact, the ratio $\alpha$ between the orbital and spin contributions is 
\begin{align}
\alpha \equiv \dfrac{\text{orbital}}{\text{spin}} =  \dfrac{4 \pi v_F^2 e \hbar t_xt_z}{3 g \mu \mu_B C_2},
\label{ratio}
\end{align}
which depends on the Fermi velocity $v_F$, the tilt, and the $g$-factor for a fixed chemical potential $\mu$,
as shown in FIG. \ref{FIG3}(b), from which we conclude that the spin contribution can be comparable
to the orbital contribution when $v_F \leq 10^{5} \mathrm{m/s}$ with $g \sim 10$.
However, for a Fermi velocity larger than $10^{5} \mathrm{m/s}$ and a small $g$-factor,
the orbital contribution for IMHE will be dominant.
Finally, we note that the ASP in momentum space shows a dipole landscape around the band crossing point,
as shown in FIG. \ref{FIG3}(c), which is different from the AOP.

\bigskip
\noindent{\textcolor{blue}{\textit{The IMHE and the chiral anomaly in WSMs.}}} ---
Although our results suggest the intrinsic IMHE can appear
even when the \textit{chiral anomaly} vanishes, a fundamental explanation is still missing.
In the semiclassical limit \cite{Berniverg2017}, the \textit{chiral anomaly} of WSM can be attributed to
the monopole charge of the conventional Berry curvature $\vect{\Omega}$. Particularly,
using the Boltzmann equation with the relaxation time approximation \cite{Spivak2016},
\begin{align*}
\partial_t \bar{f}_k + \dot{\vect{r}} \cdot \vect{\nabla} \bar{f}_k +
\dot{\vect{k}} \cdot \vect{\nabla}_{\vect{k}} \bar{f}_k = (\bar{f}_k - f_k)/\tau,
\end{align*}
along with charge density $\rho_s = \int_{\vect{k}} \mathcal{D} \bar{f}_k$ and
charge current density $\vect{j}_s = \int_{\vect{k}} \mathcal{D} \bar{f}_k \dot{\vect{r}}$,
it is easy to find 
\begin{align}
\partial_t \rho_s + \vect{\nabla} \cdot \vect{j}_s
+
\int_{\vect{k}} \mathcal{D} \dot{\vect{k}} \cdot \vect{\nabla}_{\vect{k}} \bar{f}_k 
=
\delta \rho_s/\tau.
\label{temp}
\end{align}
Performing an integration by parts on the third term of Eq. (\ref{temp}),
and using $\dot{\vect{k}}=\mathcal{D}^{-1}[-\vect{E}-\vect{v} \times \vect{B}-\vect{\Omega} (\vect{E}\cdot\vect{B})]$
from Eq. (\ref{eq1}) (ignoring the field-induced corrections)
and $\vect{\nabla}_{\vect{k}} \cdot \vect{\Omega}=s\delta(\vect{k})/2\pi$,
we arrive at the modified continuity equation \cite{Spivak, Y-Yin, D-Son}
\begin{align}
\partial_t \rho_s + \vect{\nabla} \cdot \vect{j}_s -
\frac{s}{8\pi^3} \vect{E} \cdot \vect{B} = \delta \rho_s /\tau.
\label{conservation}
\end{align}
Clearly, when $\vect{E}\cdot\vect{B} \neq 0$ in Eq. (\ref{conservation}), 
the monopole charge due to $\vect{\nabla}_{\vect{k}} \cdot \vect{\Omega}=s\delta(\vect{k})/2\pi$
gives rise to a chirality-dependent chemical potential,
which will make the current of Weyl fermions for each chirality non-conserving
and hence was dubbed the \textit{chiral anomaly}, as illustrated in Figs. \ref{FIG1}(a) and \ref{FIG1}(b).
However, for these field-induced Berry curvatures, we find that 
$\vect{\nabla}_{\vect{k}} \cdot \vect{\Omega}^B=\vect{\nabla}_{\vect{k}} \cdot \vect{\Omega}^E=0$ (see Appendix F);
namely, $\vect{\Omega}^{B/E}$ does not contribute to the monopole charge.
As a consequence, the intrinsic IMHE due to $\vect{E} \times \vect{\Omega}^B$
can be expected even when the \textit{chiral anomaly} is switched off by setting $\vect{E}\cdot\vect{B}=0$.

\bigskip
\noindent\textcolor{blue}{\textit{Discussion and summary.}}---
Recently, an in-plane Hall effect at $EB$ order was reported experimentally
in Dirac semimetal ZrTe$_5$ \cite{LiangT}. The experimental results were interpreted
using the conventional anomalous velocity $\vect{E} \times \vect{\Omega}$ and
the $\vect{E} \times \vect{\Omega}^B$, and the tilt of the Weyl cones seemed to have no effect.
Following our analytical calculations, we found that the out-of-plane tilt of WSMs is the key to capturing the intrinsic IMHE,
which in turn may explain why IMHE has been overlooked
since all of the tilt configurations explored both theoretically and experimentally
were mainly confined to the $E$-$B$ plane.
The relation between the tilt vector and the device is illustrated in FIG. \ref{FIG1}(c),
in which the tilt vector is usually along the principal axis of the crystal \cite{note12}.
Importantly, although our calculations focus on $\mathcal{P}$-related or $\mathcal{T}$-related
WPs, our conclusion can be applied to all the type-I nonmagnetic and magnetic WSMs effectively described by Eq.(\ref{WP1}),
such as the nonmagnetic WSMs TaAs family with point group $4mm$ \cite{HMWeng} and the magnetic WSM CoSn$_2$S$_2$
with point group $-3m$ \cite{Co3Sn2S2}, where the mirror symmetry must be broken \cite{globalsym}.
In addition, Dirac semimetals, such as Cd$_3$As$_2$ \cite{Cd3As2, Cd3As2exp}
and Na$_3$Bi \cite{Na3Bi, Na3Biexp}, are also potential material candidates since a Dirac cone under in-plane magnetic field
can be split into two Weyl cones. In addition,
we note that although featuring the nonlinear characteristic,
the intrinsic IMHE can, in fact, be the same order as the intrinsic anomalous Hall effect (IAHE) \cite{IAHEcompare}.

So far we have discussed only the IMHE conductivity,
but the actual experiments usually measure the resistivity.
However, the resistivity tensor can be derived from the conductivity tensor.
Particularly, we have \cite{matrix}
$\rho_{aa}=\sigma_{bb}/D$ and $\rho_{ab}=-\sigma_{ab}/D$
with $D \equiv \sigma_{aa}\sigma_{bb}-\sigma_{ab}\sigma_{ba}$,
where $a, b \in \{x, y\}$ and $a\neq b$.
In addition, $\rho_{ab}$ ($\rho_{aa}$) and $\sigma_{ab}$ ($\sigma_{aa}$)
represent the total transverse (longitudinal) resistivity and conductivity, respectively.
Moreover, $\rho_{ab}$ can be generally decomposed into four terms: 
$\rho_{ab}=\rho_{ab}^{\text{IMHE}}+\rho_{ab}^{\text{EPHE}}+\rho_{ab}^{\text{IAHE}}+\rho_{ab}^{\text{DHE}}$,
where $\rho_{ab}^{\text{IMHE}}$, $\rho_{ab}^{\text{EPHE}}$, $\rho_{ab}^{\text{IAHE}}$,
and $\rho_{ab}^{\text{DHE}}$ stand for the IMHE, extrinsic planar Hall effect (EPHE),
IAHE, and disorder-induced Hall effect (DHE) resistivities, respectively. 
Similarly, the longitudinal resistivity $\rho_{aa}$ can be generally decomposed as two terms:
$\rho_{aa}=\rho_{aa}^{\text{D}}+\rho_{aa}^{\text{EPHE}}$,
where $\rho_{aa}^{\text{D}}$ is the Drude resistivity and $\rho_{aa}^{\text{EPHE}}$ is
the longitudinal negative magnetoresistivity due to the conventional Berry curvature \cite{Nandy}.
Based on these, we discuss how to isolate the IMHE resistivity $\rho^{\text{IMHE}}_{ab}$
from the experimental result.

First of all, we note that the $\vect{B}$-independent resistivity
$\rho_{ab}^{B=0} \equiv \rho_{ab}^{\text{IAHE}}+\rho_{ab}^{\text{DHE}}$
can easily be removed by two-step measurements with and without $\vect{B}$ \cite{Nandy},
namely, $\rho_{ab}(B)=\rho_{ab}^{B\neq 0}-\rho_{ab}^{B=0}=\rho_{ab}^{\text{IMHE}}+\rho_{ab}^{\text{EPHE}}$,
where $\rho_{ab}^{B \neq 0}$ ($\rho_{ab}^{B = 0}$) stands for the measured Hall resistivity with (without) $\vect{B}$
and $\rho_{ab}(B)$ is the $\vect{B}$-dependent resistivity,
which includes the intrinsic IMHE and the extrinsic EPHE contributions. 
Fortunately, since the IMHE (EPHE) resistivity tensor is antisymmetric (symmetric) \cite{matrix},
the IMHE resistivity can be further isolated by 
$\rho_{ab}^{\text{IMHE}}
=
[\rho_{ab}(B)-\rho_{ba}(B)]/2
$.
After removing all other transverse resistivities, we find that
$\rho_{ab}^{\text{IMHE}}
=
\sigma_{ba}^{\text{IMHE}}/[
(\sigma_{aa}^{\text{D}}+\sigma_{aa}^{\text{EPHE}})
(\sigma_{bb}^{\text{D}}+\sigma_{bb}^{\text{EPHE}})
+
(\sigma_{ba}^{\text{IMHE}})^2]$,
where $\sigma_{ba}^{\text{IMHE}}=\sigma_{ba}B$,
generally feature the $2\pi$ period on $\Theta$. 
Furthermore, when $\sigma_{aa}^{\text{D}} \gg \sigma_{aa}^{\text{EPHE}}$
and $\sigma_{aa}^{\text{D}} \gg \sigma_{ba}^{\text{IMHE}}$,
we have $\rho_{ab}^{\text{IMHE}}=\sigma^{\text{IMHE}}_{ba}/(\sigma_{aa}^{\text{D}}\sigma_{bb}^{\text{D}})$;
namely, the IMHE resistivity and its conductivity have the same tilt and angular dependence.
In addition, the IMHE resistivity is also nonvanishing when the \textit{chiral anomaly} is absent,
as expected.

In summary, we predicted the intrinsic IMHE in type-I WSMs
with the extended semiclassical theory.
We found that this effect is contributed by $\vect{\Omega}^B$ (the Berry curvature induced by magnetic field)
mainly through the minimal coupling.
We revealed that an out-of-plane tilt combined with a further in-plane
tilt is the key to capturing this effect.
Different from previously reported in-plane Hall effects, the intrinsic IMHE
can survive when the \textit{chiral anomaly} is turned off,
because $\vect{\Omega}^B$ does not contribute to the monopole charge.
The experimental strategy to detect the IMHE was also discussed.
In addition to the IAHE, our work offers another intrinsic Hall transport signature for diagnosing WSMs,
and our results can be used to discuss the in-plane magnetotransport behavior contributed by
the generic three-dimensional band crossing in (topological) quantum materials.
Finally, we wish to remark that the MHE
featuring $\mathcal{P}$-even and $\mathcal{T}$-even properties
can also be expected in centrosymmetric quantum materials (such as Dirac semimetals)
even though we focused on WSMs in this work. 
As a consequence, the MHE offers a desirable tool to probe the quantum geometry
in centrosymmetric nonmagnetic quantum materials,
which may be explored in the future.

\bigskip
\noindent{{\bf Acknowledgments}} --- We thank Dr. L.Y. Wang for helpful discussions.
This work was supported by the National Natural Science Foundation of China (Grant No. 12034014).

\bigskip
\section{Appendix}
\section{Appendix A: Analytical computation of Equations (\ref{result11}) and (\ref{spinsigmayx})}
Substituting Eqs. (\ref{velocity})-(\ref{fermisurf}) into Eq. (\ref{cond11}),
we find that
\begin{align}
\sigma_{yx}^s
&=
\int_0^{+\infty} \int_0^{2\pi} \int_0^\pi
\dfrac{k^2\sin\theta dk d\theta d\phi}{(2\pi)^3}
\dfrac{\delta(k-\mu/\beta^{\pm})}{|\beta^{\pm}|}
\nonumber \\
& \times
(v_x^\pm \mathcal{F}^O_{y}-v_y^\pm \mathcal{F}^O_x)
\nonumber \\
&=
\pm \dfrac{1}{4\mu}
\int_0^{2\pi} \int_0^\pi
\dfrac{\sin\theta d\theta d\phi}{(2\pi)^3}
\dfrac{\beta^\pm}{|\beta^{\pm}|}
\nonumber \\
&\times
t_z\left[
t_x\left( \hat{k}_x^2+\hat{k}_z^2 \right) \cos\Theta 
+
t_y\left(\hat{k}_y^2+\hat{k}_z^2\right)\sin\Theta
\right]
\nonumber \\
&=
\dfrac{1}{12\pi^2\mu}
t_z
\left(
t_x\cos\Theta
+
t_y\sin\Theta
\right),
\label{sigmasyx}
\end{align}
as given by Eq. (\ref{result11}) in the main text, where $\beta^{\pm}/|\beta^{\pm}|=\pm 1$ and the identity
\begin{align}
\int_0^{2\pi} \int_0^{\pi} \hat{k}_a \sin\theta d\theta d\phi
&=
\int_0^{2\pi} \int_0^{\pi} \hat{k}_a \hat{k}_b \bar{\delta}_{ab} \sin\theta  d\theta d\phi
\nonumber \\
&=
\int_0^{2\pi} \int_0^{\pi} \hat{k}_a \hat{k}_b \hat{k}_c \sin\theta d\theta d\phi = 0
\label{identity}
\end{align}
was used, where $\bar{\delta}_{ab}=1-\delta_{ab}$. Similarly, by substituting Eq. (\ref{FSxy})
into Eq. (\ref{cond11}), we arrive at:
\begin{align}
\sigma_{yx}^s
&= 
\pm g \mu_B \int_0^{+\infty} \int_0^{\pi} \int_0^{2\pi} 
\dfrac{k^2\sin\theta dk d\theta d\phi}{(2\pi)^3}
\nonumber \\
& \times
\dfrac{\delta(k-\mu/\beta^{\pm})}{|\beta^{\pm}|}
\dfrac{\hat{k}_z}{2k^2}
\left( v_x^{\pm} \cos\Theta + v_y^{\pm} \sin\Theta \right)
\nonumber \\
&=
\dfrac{g \mu_B}{16 \pi^3} \left(C_1 \cos\Theta+C_2 \sin\Theta \right),
\label{spinsigmayxhalf}
\end{align}
where $C_i \equiv \pm \int_0^{\pi}\int_0^{2\pi} \lambda_i^\pm (\vect{t};\theta,\phi)d\theta d\phi$ is a dimensionless constant
with
\begin{align}
\lambda_1^\pm &= (t_x \pm \sin\theta\cos\phi)\cos\theta\sin\theta/|\beta^{\pm}|,
\label{lambda1}
\\
\lambda_2^\pm &= (t_y \pm \sin\theta\sin\phi)\cos\theta\sin\theta/|\beta^{\pm}|.
\label{lambda2}
\end{align}
By duplicating Eq. (\ref{spinsigmayxhalf}), we obtain Eq. (\ref{spinsigmayx}) in the main text.

\smallskip
\subsection{Appendix B: The contribution from the second term of Equation (\ref{orbital})}
From the second term of Eq. (\ref{orbital}), the AOPs contributed by the quantum metric dipole
for Eq. (\ref{WP1}) are given by:
\begin{align}
\mathcal{F}^O_x=-\dfrac{\hat{k}_z\sin\Theta}{8k^3},
\quad
\mathcal{F}^O_y=-\dfrac{\hat{k}_z\cos\Theta}{8k^3}.
\end{align}
Substituting these two expressions into Eq.(\ref{cond11}) and using Eq.(\ref{identity}),
we immediately obtain $\sigma^s_{yx}=0$.

\smallskip
\subsection{Appendix C: The Berry curvature contribution}
Combining the conventional Berry curvature with the first-order wavepacket energy correction under magnetic field,
we find \cite{GaoY2014PRL, XiaoC2022arxiv}
\begin{align}
\sigma^{s}_{yx} \equiv -\int_{\vect{k}} f'(\epsilon^{\pm}-\mu) \epsilon^{(B,\pm)} \Omega_z^{\pm},
\end{align}
where $\epsilon^{(B,\pm)} \equiv \vect{B}\cdot\vect{m}/B$,
with $\vect{m}$ being the orbital magnetic moment \cite{D-Xiao}.
At zero temperature, using Eq. (\ref{identity}) we find
\begin{align}
\sigma^{s}_{yx} 
=
&-
\int_0^{+\infty}
\int_0^{2\pi}
\int_0^{\pi}
\dfrac{k^2\sin\theta dk d\theta d\phi}{(2\pi)^3}
\dfrac{\delta(k-\mu/\beta^\pm) }{|\beta^\pm|}
\nonumber \\
& \times
\left[ \mp \dfrac{\hat{k}_z(\hat{k}_x\cos\Theta+\hat{k}_y\sin\Theta)}{4k^3} \right]
=0.
\end{align}

\smallskip
\subsection{Appendix D: The chiral velocity contribution}
By the chiral velocity $\vect{B}(\bar{\vect{v}}\cdot\bar{\vect{\Omega}})$,
a current at $EB$ order for Eq. (\ref{WP1}) is found to be
\begin{align*}
\vect{j}^s
&\equiv
\vect{B} \int_{\vect{k}} f (\vect{v}\cdot\vect{\Omega}^E)
=
E \vect{B}
\int_{\vect{k}} f  \left(\mp t_y\dfrac{k_z}{2k^5} \pm t_z \dfrac{k_y}{2k^5} \right)
\\
&=
E \vect{B} \int_{\vect{k}} f
\left[\mp t_y \partial_z \left(-\dfrac{1}{6k^3}\right) 
\pm
t_z\partial_y\left( -\dfrac{1}{6k^3}\right) \right]
\\
&=
\dfrac{E\vect{B}}{6}
\int_0^{+\infty} \int_0^{2\pi} \int_0^{\pi}
\dfrac{k^2\sin\theta dk d\theta d\phi}{(2\pi)^3}
\nonumber \\
&
\times
\dfrac{\delta(k-\mu/\beta^{\pm})}{|\beta^\pm|} \dfrac{\left(\mp t_y v_z^{\pm} \pm t_zv_y^{\pm} \right)}{k^3}
\\
&=
\dfrac{E\vect{B}}{6\mu}
\int_0^{2\pi} \int_0^{\pi}
\dfrac{\beta^{\pm}}{|\beta^\pm|}
\left(\mp t_y v_z^{\pm} \pm t_zv_y^{\pm} \right)
\dfrac{\sin\theta d\theta d\phi}{(2\pi)^3}
=0.
\end{align*}

\subsection{Appendix E: The dispersive velocity contribution}
The energy correction at $EB$ order is given by \cite{XiaoCPRB2021}
\begin{align*}
\epsilon_{EB}^{\pm} 
\equiv
B_a\epsilon_{abc}\mathcal{A}^{(E,\pm)}_b v_c^{\pm}
=
\dfrac{\left(\eta_{1}^\pm \cos\Theta + \eta_{2}^\pm \sin\Theta \right)EB}{4k^3},
\end{align*}
where
$\eta_1^\pm = \mp \hat{k}_x(t_z\hat{k}_y-t_y\hat{k}_z)$
and
$\eta_2^\pm = \mp t_x \hat{k}_x \hat{k}_z \mp t_z(\hat{k}_y^2+\hat{k}_z^2)-\hat{k}_z$
for $\vect{E}=(E,0,0)$ and $\vect{B}=B(\cos\Theta,\sin\Theta,0)$;
for $\vect{E}=(0,E,0)$ and $\vect{B}=B(\cos\Theta,\sin\Theta,0)$, we have
$\eta_1^\pm = \pm t_y \hat{k}_y \hat{k}_z \pm t_z(\hat{k}_x^2+\hat{k}_z^2) + \hat{k}_z$
and
$\eta_2^\pm = \pm \hat{k}_y(t_z\hat{k}_x-t_x\hat{k}_z)$. 
By inserting these expressions into $j_a^s=\int_{\vect{k}} f \partial_a \epsilon_{EB}^\pm \equiv \bar{\sigma}_{ab}^s E_bB$,
we find
\begin{align*}
\bar{\sigma}_{yx}^s = \dfrac{1}{12\pi^2\mu} t_y t_z \sin\Theta,
\quad
\bar{\sigma}_{xy}^s = -\dfrac{1}{12\pi^2\mu} t_x t_z \cos\Theta.
\end{align*}
Furthermore, by antisymmetrizing this result, we finally arrive at:
\begin{align}
\sigma_{yx}^s = \dfrac{\bar{\sigma}_{yx}^s-\bar{\sigma}_{xy}^s}{2}
=
\dfrac{1}{24\pi^2\mu}t_z(t_x\cos\Theta+t_y\sin\Theta),
\end{align}
which displays the same behavior as Eq. (\ref{result11}) in the main text.

\smallskip
\subsection{Appendix F: $\vect{\Omega}^{B/E}$ and monopole charge}
For $\vect{B}=(B_x, B_y, B_z)$, it is easy to show that
$\vect{\Omega}^B$ calculated from Eq. (\ref{WP1}) satisfies 
\begin{align}
\nabla \cdot \vect{\Omega}^B=0 \qquad (k \neq 0).
\label{divB}
\end{align}
Next we consider $\nabla \cdot \vect{\Omega}^B$ at the singular point $k=0$.
Using the divergence theorem, we have
\begin{align}
\int_V \nabla \cdot \vect{\Omega}^B dV
=
\oint_S \vect{\Omega}^B \cdot \hat{\vect{n}} dS
\end{align}
where $\hat{\vect{n}}$ is the normal vector for the surface $S$ enclosing the volume $V$.
Using Eq. (\ref{divB}), we can choose a tiny sphere $V_1$ to enclose the singularity point $k=0$,
then
\begin{align}
\int_{V_1} \nabla \cdot \vect{\Omega}^B dV_1
=
\oint_{S_1} \vect{\Omega}^B \cdot \hat{\vect{n}} dS_1
\end{align}
where $S_1$ is defined as $k_x^2+k_y^2+k_z^2=r_\epsilon^2$ with $r_\epsilon \rightarrow 0$
and the normal vector is explicitly given by 
$\hat{\vect{n}}=(k_x,k_y,k_z)/k$.
Performing the surface integral, we find:
\begin{align*}
&\oint_{S_1} \vect{\Omega}^B \cdot \hat{\vect{n}} dS_1 
=
\int_0^{\pi} d\theta \int_0^{2\pi} d\phi 
r_\epsilon^2 \sin\theta
\nonumber \\
& \times
\dfrac{B_x r_\epsilon \sin\theta \cos\phi+B_yr_\epsilon\sin\theta\sin\phi+B_zr_\epsilon\cos\theta}{4r_\epsilon^5}
=0,
\end{align*}
which indicates $\nabla \cdot \vect{\Omega}^B=0$ for $k=0$. 
This is, in fact, the case for $\vect{\Omega}^E$.
We conclude that $\vect{\Omega}^B$ does not contribute to the monopole charge due to
the conventional Berry curvature $\vect{\Omega}$ of WSMs,
namely, $\nabla \cdot \bar{\vect{\Omega}} = \nabla \cdot \vect{\Omega} =2\pi\delta(\vect{k})$.


\begin{thebibliography}{00}
\bibitem{comma}
The comma in MHE conductivity indicates that $b$ and $c$ are not symmetric like the ENHE conductivity.
\bibitem{inplane}
Here "in-plane" means that the Hall current $\vect{j}$, the driving electric field $\vect{E}$,
and the driving magnetic field $\vect{B}$ share the same plane.
\bibitem{Esum}
Here the Einstein summation convention for repeated index is assumed.
\bibitem{note12} For instance, the first-principles calculation
shows that the Weyl cone is tilted along the $b$-axis for WTe$_2$\cite{X-Dai, C-Wang1}.
\bibitem{globalsym}
Following Neumann's principle implemented in the online Bilbao database \cite{Bilbao},
by defining the Jahn notation $e\{V^2\}V$ for $\sigma_{ab,c}$, we find that
both $4mm$ and $-3m$ can not support $\sigma_{ab,c}$ when the mirror symmetries exist.
Note that the Jahn notation $e\{V^2\}V$ for $\sigma_{ab,c}$ can also be used to
identify when $\sigma_{ab,c}$ will be supported in other WSMs.
\bibitem{matrix}
In matrix form, we have
$
\begin{pmatrix} \rho_{xx} & \rho_{xy} \\ \rho_{yx} & \rho_{yy} \end{pmatrix}
=
\begin{pmatrix} \sigma_{xx} & \sigma_{xy} \\ \sigma_{yx} & \sigma_{yy} \end{pmatrix}^{-1}
=
\frac{1}{\sigma_{xx}\sigma_{yy}-\sigma_{xy}\sigma_{yx}}
\begin{pmatrix}    \sigma_{yy}        & -\sigma_{xy} \\ -\sigma_{yx} &   \sigma_{xx}   \end{pmatrix}
$,
from which we obtain $\rho_{xx}=\sigma_{yy}/D$, $\rho_{yy}=\sigma_{xx}/D$, $\rho_{xy}=-\sigma_{xy}/D$,
and $\rho_{yx}=-\sigma_{yx}/D$ with $D=\sigma_{xx}\sigma_{yy}-\sigma_{xy}\sigma_{yx}$.
Importantly, the transverse component of conductivity always features the same symmetry as
the transverse component of resistivity, namely, $\sigma_{xy}=\pm\sigma_{yx} \Leftrightarrow \rho_{xy}=\pm \rho_{yx}$.
\end{thebibliography}
\end{document}